\documentclass[preprint]{aastex}

\usepackage{times}
\usepackage{wasysym}
\usepackage{rotating}  

\newcommand{\degree}{\mbox{\ensuremath{^\circ}}}   
\newcommand{\teff}{\mbox{$T_{\rm eff}$}}
\newcommand{\logg}{\mbox{$\log g$}}
\newcommand{\vsini}{\mbox{$v \sin i$}}
\newcommand{\mictrb}{\mbox{$v_{\rm mic}$}}
\newcommand{\mactrb}{\mbox{$v_{\rm mac}$}}

\newcommand{\kms}{\mbox{km\,s$^{-1}$}}
\newcommand{\halpha}{\mbox{$H_\alpha$}}
\newcommand{\vmag}{\mbox{V$_{\rm mag}$}}

\newcommand{\rhostar}{\ensuremath{\rho_\star}}
\newcommand{\rhosun}{\ensuremath{\rho_\odot}}
\newcommand{\rhoj}{\ensuremath{\rho_{\rm J}}}
\newcommand{\rhopl}{\ensuremath{\rho_{p}}}
\newcommand{\rj}{R\ensuremath{_{\rm J}}}
\newcommand{\mj}{M\ensuremath{_{\rm J}}}
\newcommand{\rsun}{R\ensuremath{_\odot}}
\newcommand{\msun}{M\ensuremath{_\odot}}
\newcommand{\rpl}{\ensuremath{R_{p}}}
\newcommand{\mpl}{\ensuremath{M_{p}}}
\newcommand{\rstar}{\ensuremath{R_\star}}
\newcommand{\mstar}{\ensuremath{M_\star}}

\shorttitle{Independent discovery of HAT-P-14b}
 \shortauthors{E. K. Simpson et al.}
 
\begin{document}
 
\title{Independent discovery and refined parameters of the transiting exoplanet HAT-P-14\lowercase{b}}

\author{E. K.~Simpson,$^1$ S. C. C.~Barros,$^1$  D. J. A.~Brown,$^2$ A.~Collier Cameron,$^2$ D.~Pollacco,$^1$ I.~Skillen,$^3$ H. C.~Stempels,$^4$ I.~Boisse,$^5$ F.~Faedi,$^1$ G.~H\'{e}brard,$^5$ J.~McCormac,$^1$  P.~Sorensen,$^6$  R. A.~Street,$^7$ J.~Bento,$^{8}$  F.~Bouchy,$^{5,9}$ O. W.~Butters,$^{10}$ B.~Enoch,$^2$ C. A.~Haswell,$^{11}$  L.~Hebb,$^{12}$ S. Holmes,$^{11}$  K.~Horne,$^2$ F. P.~Keenan,$^1$ T. A.~Lister,$^7$ G. R. M.~Miller,$^2$ V.~Moulds,$^1$ C. Moutou,$^{13}$ A. J.~Norton,$^{11}$ N.~Parley,$^2$ A. Santerne,$^{13}$  I.~Todd,$^1$ C. A.~Watson,$^1$ R. G.~West,$^{10}$ and P. J.~Wheatley$^{8}$}
\affil{
$^1$ Astrophysics Research Centre, School of Mathematics and Physics, Queen's University Belfast, Belfast, BT7 1NN, UK \\
$^2$ School of Physics and Astronomy, University of St Andrews, North Haugh, St Andrews, Fife KY16 9SS, UK\\
$^3$ Isaac Newton Group of Telescopes, Apartado de Correos 321, E-38700 Santa Cruz de la Palma, Tenerife, Spain \\
$^4$ Department of Physics and Astronomy, Uppsala University, Box 516, SE-751 20 Uppsala, Sweden\\
$^5$ Institut d'Astrophysique de Paris, UMR7095 CNRS, Universit\'e Pierre \& Marie Curie, 98bis Bd Arago, 75014 Paris, France\\
$^6$ Nordic Optical Telescope, Apartado de Correos 474, E-387 00 Santa Cruz de la Palma, Canary Islands, Spain\\
$^7$ Las Cumbres Observatory Global Telescope Network, 6740 Cortona Drive Suite 102, Goleta, CA 93117, USA\\
$^8$ Department of Physics, University of Warwick, Coventry CV4 7AL, UK\\
$^9$ Observatoire de Haute-Provence, CNRS/OAMP, 04870 St Michel l'Observatoire, France\\
$^{10}$ Department of Physics and Astronomy, University of Leicester, Leicester, LE1 7RH \\
$^{11}$ Department of Physics and Astronomy, The Open University, Milton Keynes, MK7 6AA, UK\\
$^{12}$ Department of Physics and Astronomy, Vanderbilt University, Nashville, TN 37235, USA\\
$^{13}$ Laboratoire d'Astrophysique de Marseille, 38 rue Fr\'ed\'eric Joliot-Curie, 13388 Marseille cedex 13, France \\
}

\begin{abstract}
We present SuperWASP observations of HAT-P-14b, a hot Jupiter discovered by Torres et al. The planet was found independently by the SuperWASP team and named WASP-27b after follow-up observations had secured the discovery, but prior to the publication by Torres et al. Our analysis of HAT-P-14/WASP-27 is in good agreement with the values found by Torres et al. and we refine the parameters by combining our datasets. We also provide additional evidence against astronomical false positives. Due to the brightness of the host star, $V_{\rm mag}$ = 10, HAT-P-14 is an attractive candidate for further characterisation observations. The planet has a high impact parameter, b = 0.907 $\pm$ 0.004, and the primary transit is close to grazing. This could readily reveal small deviations in the orbital parameters indicating the presence of a third body in the system, which may be causing the small but significant orbital eccentricity, $e$ = 0.095 $\pm$ 0.011. The system geometry suggests that the planet narrowly fails to undergo a secondary eclipse. However, even a non-detection would tightly constrain the system parameters. 

\end{abstract}


\section{Introduction}\label{Intro}

There has been a rapid increase in the number of transiting planets discovered each year due to dedicated ground-- and space-- based surveys: HAT \citep{Bakos02}, TrES \citep{Alonso04}, XO \citep{McCullough05}, WASP \citep{Pollacco06}, CoRoT \citep{Baglin06} and Kepler \citep{Kepler}. This trend looks set to continue, with the discovery of over 35 new planets published already this year (mid 2010), which represents more than a third of the total number of transiting planets known. We have now surpassed one hundred transiting planet discoveries. By increasing the number of well-characterised transiting planets, we are beginning to be able to test theories of planet formation and evolution on a secure statistical basis. For example, the obliquity of hot Jupiters appears to show that the mechanism by which planets migrate cannot be explained solely by planet-disc interactions \citep{FW09, Triaud10}. 

Transiting planets orbiting the brightest stars are extremely valuable due to the extensive follow-up observations which can be undertaken. This includes secondary eclipse observations to determine atmospheric properties and refine orbital parameters. In order to find planets orbiting bright stars, a wide field-of-view must be employed to encompass enough stars to make the detection probability significant. It is therefore inevitable that some areas of the sky will be monitored by multiple groups, and indeed several transiting planets have been discovered independently: WASP-11b \citep{West09} = HAT-P-10b \citep{Bakos09} and XO-5 \citep{Burke08,Pal09}.

Here we describe observations of the transiting planet HAT-P-14b \citep[][hereafter T10]{Torres10planet} by the SuperWASP survey. The star was monitored by the SuperWASP station on La Palma and followed up with photometric and spectroscopic observations using the Faulkes Telescope North, Liverpool Telescope, Observatoire de Haute-Province and Nordic Optical Telescope. The planet was named WASP-27b after the detection was secured, prior to the publication of the T10 paper. While the HAT-P-14b designation is to be preferred in recognition of priority of publication, it is important that we retain WASP-27b as a secondary designation for statistical completeness in studies of the physical properties of independently-discovered WASP planets.  

Section \ref{OandM} describes the observations of HAT-P-14b/WASP27b, including the SuperWASP discovery data and photometric and spectroscopic follow-up. The results of the derived system parameters are presented in Section \ref{Results}, including an independent spectral analysis and a joint fit with the T10 data in order to refine the planetary parameters. We discuss our findings and compare them to those found by T10 in Section \ref{Conc}. 

\section{Observations} \label{OandM}

\subsection{SuperWASP photometry}

 \begin{figure}
 \centering
\includegraphics[width=10cm]{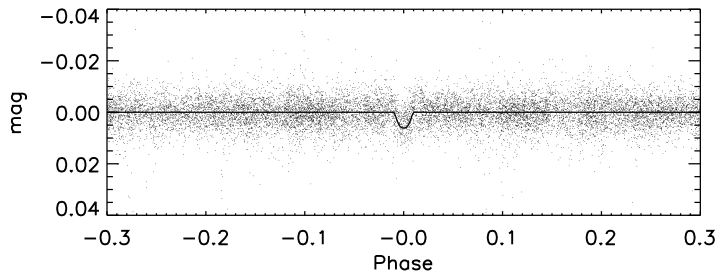} 
\caption{The combined, unbinned SuperWASP light curve for HAT-P-14, folded on the orbital period of P =  4.627654 d. Superimposed is the model transit light curve, based on the system parameters determined from a global fit, see Section \ref{Planet}.}
\label{SWASPlc} 
\end{figure}

HAT-P-14 is a bright, (\vmag\ = 9.98) star in the constellation Virgo, located at $\mathrm {\alpha_{J2000} = 17^{h}20^{m}27^{s}.87}$, $\mathrm{\delta_{J2000} = +38\degr14\arcmin31\arcsec.9}$ (GSC 3086-00152; 2MASS 17202788+3814317). It has been observed by SuperWASP (La Palma) since 2004 and a total of 25,474 photometric points have been collected.  A periodic signature was detected in the observations obtained between March and August in 2007, 2008 and 2009.  All the data were processed using the pipeline described in \citet{Cameron06}, yielding a detection of a transit-like feature with a period of 4.6278 d in multiple cameras and successive seasons. The folded light curve is shown in Figure \ref{SWASPlc}. 

The target underwent several consistency tests aimed at eliminating false positives (see \citealt{Cameron07}). Resolved blends were ruled out by inspecting Digitized Sky Survey (DSS) images which showed the star to be isolated within the WASP photometric aperture. No significant ellipsoidal variability was measurable in the WASP light curve folded on the transit period. The transit depth and duration yielded a planet-like radius for the companion, and a stellar density appropriate to a main-sequence host star of the effective temperature derived from the 2MASS colours. Having passed all these tests, the star was selected for follow-up observations.

\subsection{Photometric follow-up}

\begin{figure} 
\centering
\includegraphics[height=\linewidth]{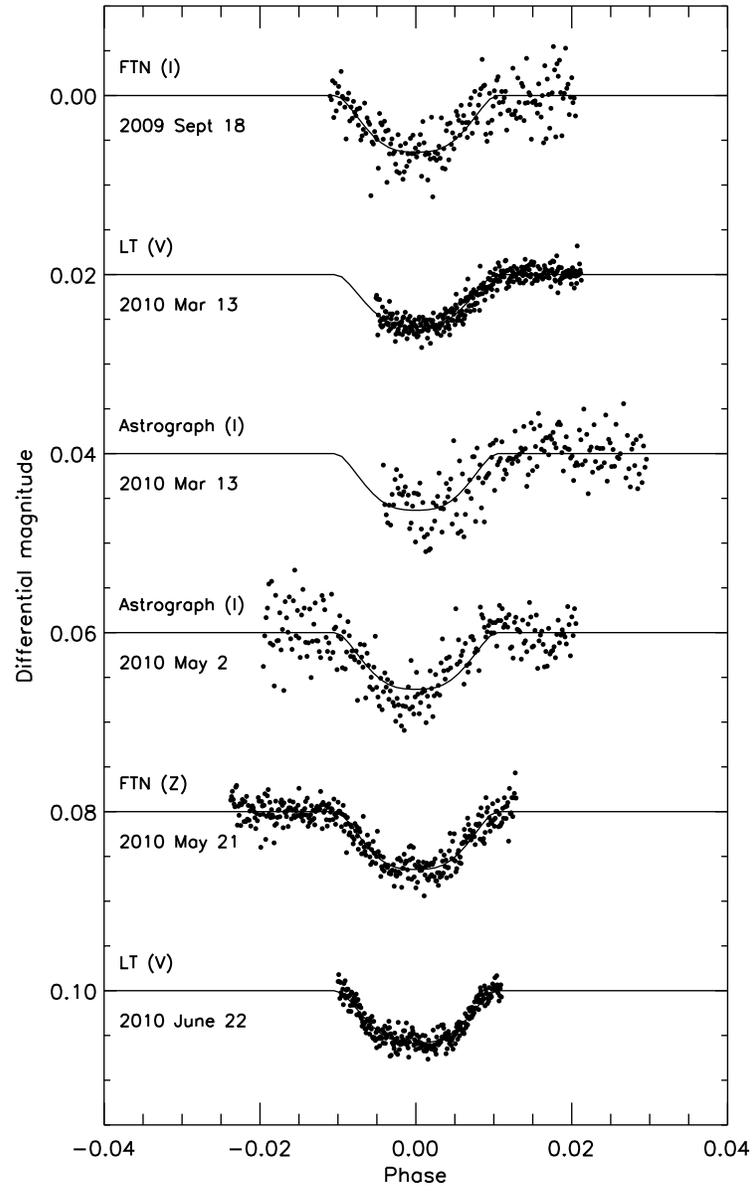} 
\caption{Photometry of transit events of HAT-P-14b taken with FTN, Astrograph and LT in various bands. The light curves have been offset from zero by arbitrary amounts for clarity. Superimposed are the model transit light curves based on the determined system parameters (see Section \ref{Planet}).}
\label{lcs} 
  \end{figure} 

We obtained six further light curves in order to refine the photometric parameters, shown in Figure \ref{lcs}. All photometric data presented in this paper are available from the NStED database.\footnote{http://nsted.ipac.caltech.edu}

Two full transits were observed with Faulkes-North (FTN, Hawaii), on 2009 September 18 (SDSS i-band) and 2010 May 21 (Pan-STARRS-z filter). The first transit was taken using the Merope camera (EM01) in binning 2$\times$2 mode and defocussed to enable an exposure time of 40s.  Aperture photometry was performed using an 18-pixel radius aperture and ensemble photometry using the 2 bright, non-variable comparison stars within the field-of-view. By 2010 May 21, the new Spectral camera (FS03) was available at FTN and defocussed to obtain 20s exposures. The 10.36$\times$10.22 arc-minute field-of-view included 4 bright comparison stars. These observations took advantage of the 4ag autoguider to maintain the locations of the stars to within an RMS $\sim$ 2 pixel of the same location throughout the observations and led to a noticable improvement in the photometric precision. All the FTN data were reduced using the pipeline written at Liverpool John Moores University, and photometry performed using the IRAF/DAOPHOT package.

A partial transit on 2010 March 13 was observed simultaneously by the robotic  2.0-m Liverpool Telescope (LT) and an 18cm Takahashi Astrograph telescope, both located on La Palma. Two further full transits were observed, on 2010 May 2 using the Astrograph, and on 2010 June 22 by the LT. 

We used an Andor 1$\times$1K e2v CCD detector with a 5 s readout time mounted on the Astrograph. For the first transit, the telescope was defocussed with FWHM = 20\arcsec and 685 exposures of duration 20 s were obtained in the 4.8 hour period. For the second transit, the exposure time was increased to 30 s and FWHM = 28\arcsec giving 456 observations in the 4.5 hour period. The data were reduced and differential photometry performed using the PYRAF/DAOPHOT package. The observations were subsequently binned with 3 and 2 points per bin, respectively.

We also used the RISE frame transfer CCD located on the LT \citep{Gibson08, Steele08}. For both transit observations, the telescope was defocused by -1.2mm giving FWHM = 18 and 21 pixels on the respective nights (0.54 arcsec/pixel). The CCD was binned 2x2 and exposures of 3.7 s taken with effectively no dead time. During the 3.0 h and 2.3 h periods, 3520 and 3290 frames were taken, respectively. The data were reduced and differential photometry performed using the ULTRACAM pipeline \citep{Dhillon07}, and then binned with 10 points per bin.

\subsection{Spectroscopic Observations}

\begin{table} 
\centering
\caption{Radial velocities (RV) and line bisector spans ($V_{\rm span}$ ) of HAT-P-14.}
\begin{tabular}{cccrc}
\hline \hline
BJD		&RV		&Error	& $V_{span}$ & Instrument \\
-2 450 000.0	&(\kms)	&(\kms)	 & (\kms) & \\
\hline
5024.5363   &     -20.6620  &    0.0071          & -0.0031 & FIES\\
5025.5236   &     -20.4951  &    0.0104          &  0.0030 & FIES\\
5026.4324   &     -20.2582  &	0.0105  	& -0.0155 & FIES\\
5026.6481   &     -20.2427  &	0.0145  	&  0.0010 & FIES\\
5040.5041   &     -20.1997  &    0.0179          & -0.0326 & FIES\\
5041.5629   &     -20.3024  &    0.0178          & -0.0260 & FIES\\
5085.3553   &     -20.5455  &    0.0072  	&  0.0119 & FIES\\
5086.3534   &     -20.2943   &   0.0150          & -0.0064 & FIES\\
5087.3509   & 	 -20.2642 &	0.0126  	& -0.0072 & FIES\\
5096.4062   & 	 -20.2160 &	0.0147  	&  0.0046 & FIES\\
5098.3406   & 	 -20.6554 &	0.0073  	& -0.0336 & FIES\\
5099.3361   & 	 -20.5338 &	0.0088  	& -0.0087 & FIES\\
 5100.3558   & 	 -20.2565 &	0.0127  	& -0.0021 & FIES\\
5101.3677   & 	 -20.2778 &	0.0121  	& -0.0029 & FIES\\
5119.3355   & 	 -20.2386 &	0.0150  	& -0.0014 & FIES\\
5291.6152   & 	 -20.3542 &	0.0094  	& -0.0026 & FIES\\
5291.6300   & 	 -20.3725 &	0.0080  	&  0.0090 & FIES\\
5291.6449   & 	 -20.3932 &	0.0075  	&  0.0016 & FIES\\
5291.6597   & 	 -20.3904 &	0.0095  	& -0.0074 & FIES\\
5291.6746   & 	 -20.3904 &	0.0063  	& -0.0294 & FIES\\
5291.6895   & 	 -20.3837 &	0.0079  	& -0.0231 & FIES\\
5291.7043   & 	 -20.4016 &	0.0086  	& -0.0029 & FIES\\
5291.7192   & 	 -20.3857 &	0.0063  	& -0.0054 & FIES\\
5291.7340   & 	 -20.3984 &	0.0065  	&  0.0062 & FIES\\
5068.3272 & -20.1678     &       0.0146 &  0.0236   & SOPHIE \\
5071.4331 & -20.5014	&0.0123&   0.0540  & SOPHIE \\
5072.3251 & -20.2567	&0.0126& -0.0163  & SOPHIE \\
5073.3273 & -20.1498	&0.0166&  -0.0242   & SOPHIE \\
5074.4166 & -20.4524	&0.0133&   0.0372   & SOPHIE \\
5075.4225 & -20.6275	&0.0132 &  0.0031   & SOPHIE \\
5076.3691 &  -20.4281	&0.0192 &  -0.0165   & SOPHIE \\
5079.3532 &  -20.5479	&0.0144 &  0.0437   & SOPHIE \\
5080.3964 & -20.5847	&0.0162  &  0.0308   & SOPHIE \\

\hline
\end{tabular}
\label{rv-data}
\end{table}

\begin{figure} 
\centering
\includegraphics[width=10cm]{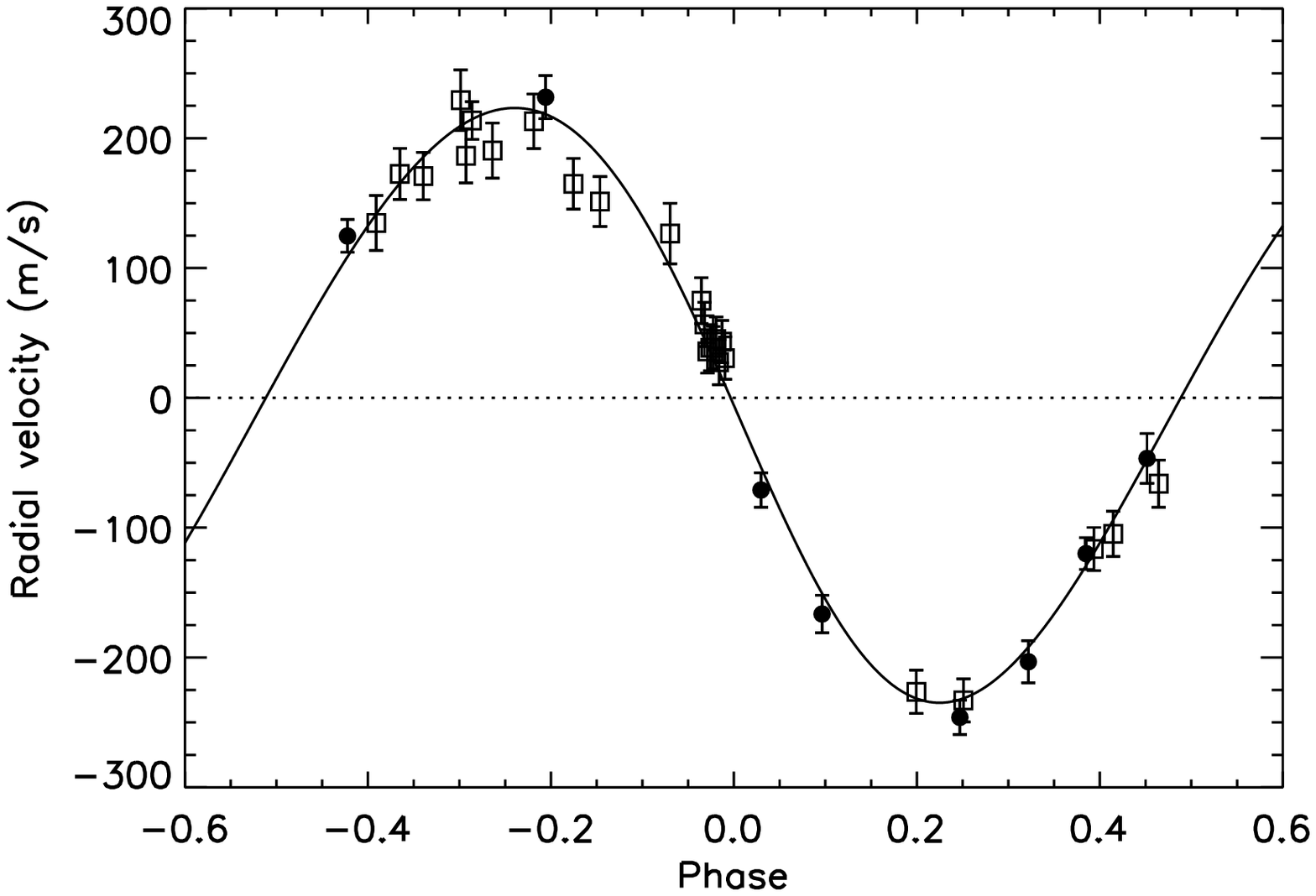} 
\includegraphics[width=10cm]{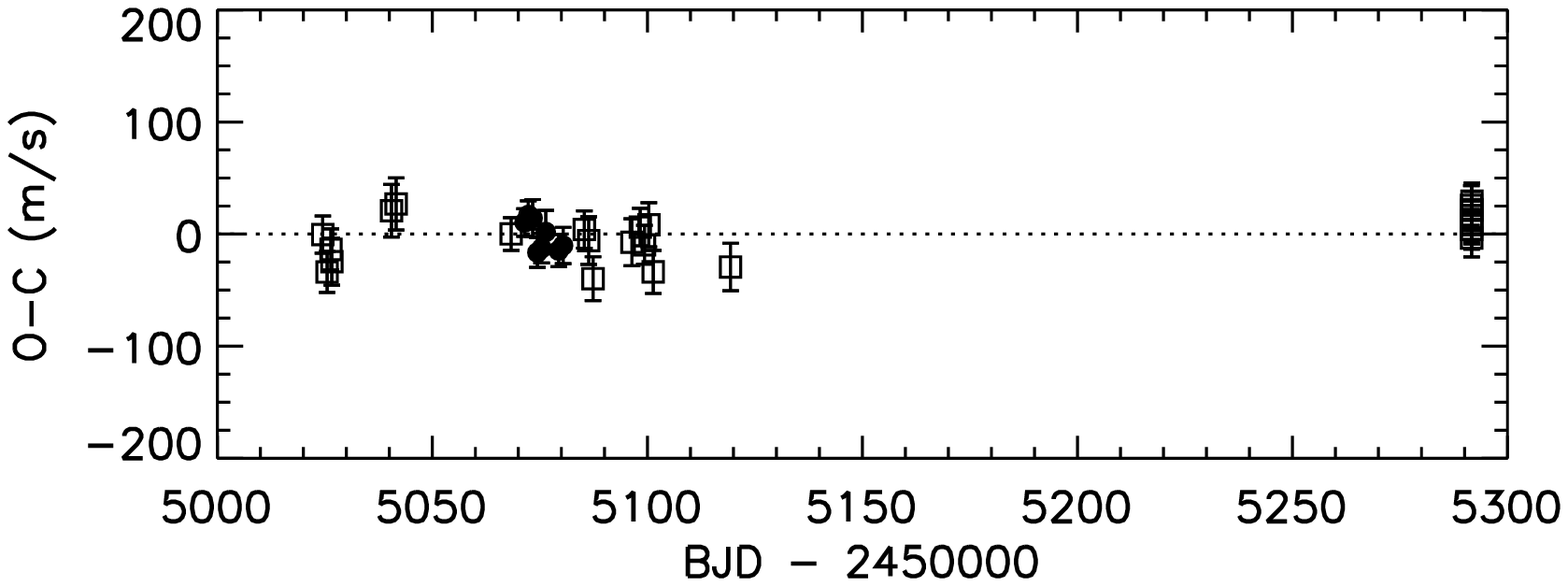} 
\includegraphics[width=10cm]{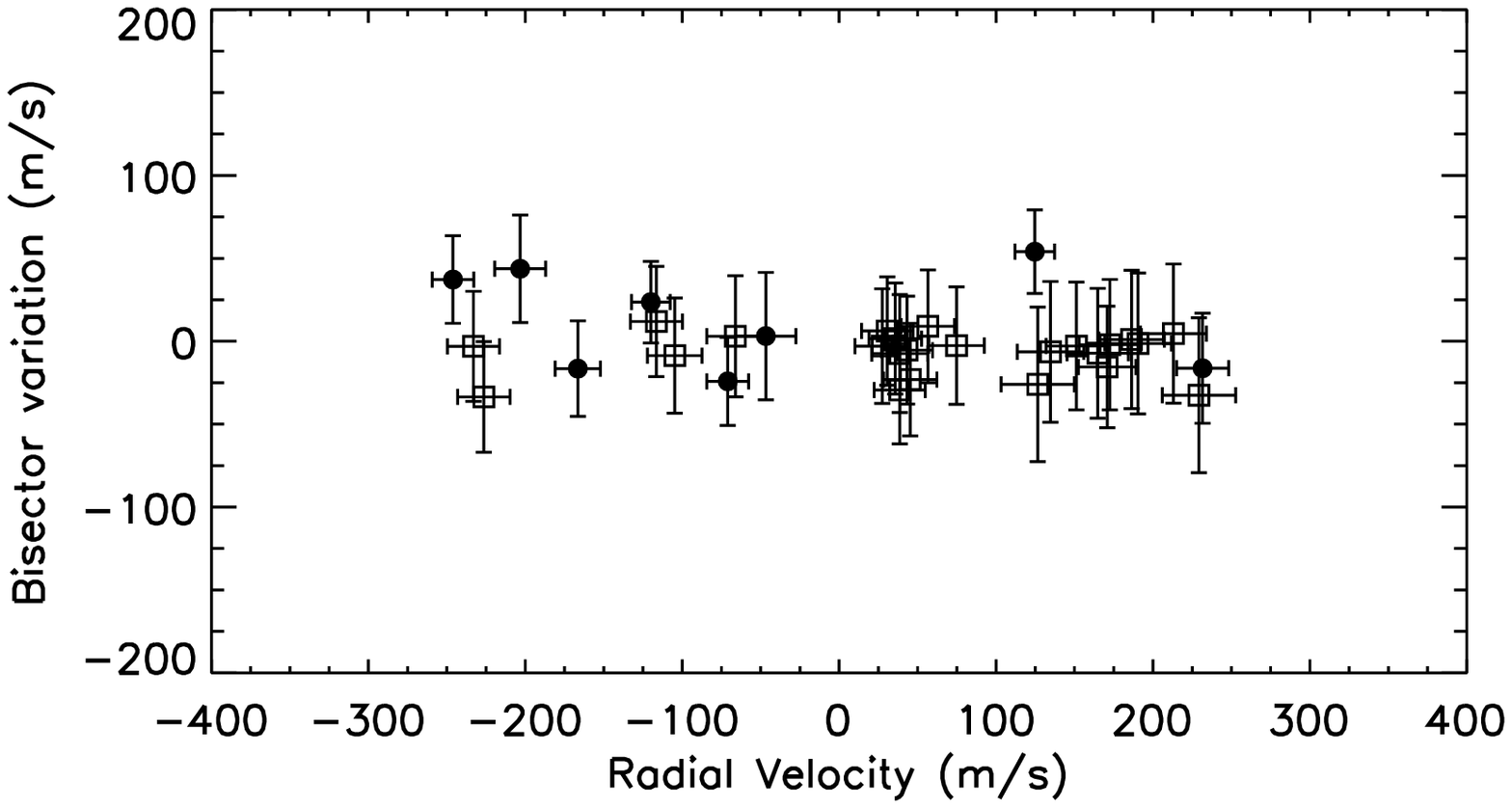} 
\caption{Upper panel: Phase folded radial velocity measurements of HAT-P-14, combining data from NOT/FIES (open squares) and OHP/SOPHIE (filled circles), and superimposed with the best-fit model RV curve based on the determined system parameters. The systematic velocity for each dataset was subtracted from the RVs. Middle panel: Residuals from the orbital fit plotted against time. No long-term trend is visible. Lower panel: The bisector span measurements as a function of radial velocity, showing no correlation. The uncertainties in the bisector spans was taken as twice the RV uncertainties. } 
\label{RV} 
\end{figure} 

We obtained follow-up spectroscopic observations to determine the planetary, orbital and stellar parameters. The star was initially observed using the FIbre-fed Echelle Spectrograph (FIES) mounted on the 2.5-m Nordic Optical Telescope. In total, 24 usable spectra  were obtained between 2009 July 12 and 2010 April 05. Two spectra were removed from the analysis due to Moon contamination affecting the radial velocity determination. FIES was used in medium resolution mode ($R$ = 46,000) with simultaneous ThAr calibration, and observations were conducted using exposure times of 1200s.  We used the bespoke data reduction package FIEStool\footnote{http://www.not.iac.es/instruments/fies/fiestool/FIEStool.html} to extract the spectra.  An IDL cross- correlation routine was used to obtain radial velocities (RVs) by fitting gaussians to the cross-correlation functions (CCFs) of 30 spectral orders and taking the mean. A template spectrum was constructed by shifting and co-adding the spectra, against which the individual spectra were cross-correlated to obtain the final velocities. The template was cross-correlated with a high signal-to-noise spectrum of the Sun to obtain the absolute velocity to which the relative RVs were shifted. The RV uncertainty is given by $\sigma$ = RMS$(v)$ /  $\sqrt{N}$,  where $v$ is the RV of the individual orders and $N$ is the number of orders. Line bisector spans ($V_{\rm span}$) were computed using the difference in the position of the midpoint of the CCF at 25\% and 75\% of its depth. 

A further nine spectra were taken between 2009 August 24 and 2009 September 05 with the stabilised echelle spectrograph SOPHIE at the 1.93-m telescope of Observatiore de Haute-Provence \citep{Perruchot08,Bouchy09}. The spectrograph was used in high efficiency mode (resolution $R$ = 40,000) and the observations were taken with a signal-to-noise ratio S/N$\sim$30 to minimise the Charge Transfer Inefficiency (CTI) effect \citep{Bouchy09}. Two 3 arc-second diameter optical fibres were used, the first centred on the target and the second on the sky to simultaneously measure the background to remove contamination from scattered moonlight. The spectra were extracted using the SOPHIE pipeline \citep{Perruchot08} and RVs computed from a weighted cross correlation of each spectrum with a numerical mask of G2 spectral type, as described by \citet{Baranne96} and \citet{Pepe02}. Cross correlation with other masks (F0, K5 and M5) produced an identical RV semi-amplitude, indicating that the variation is unlikely to be caused by a blended eclipsing binary system of unequal masses. 
 
The radial velocities and line bisector spans  are shown in Table \ref{rv-data} and plotted in Figure \ref{RV}. A possible correlation between the bisector span and radial velocity was noted by T10 but was attributed to moonlight contamination rather than to a blended eclipsing binary. We find no correlation, supporting the signal's origin as a planetary companion rather than an astronomical false positive.

\subsection{High resolution imaging}

\begin{figure}	
\centering
\includegraphics[width=8cm]{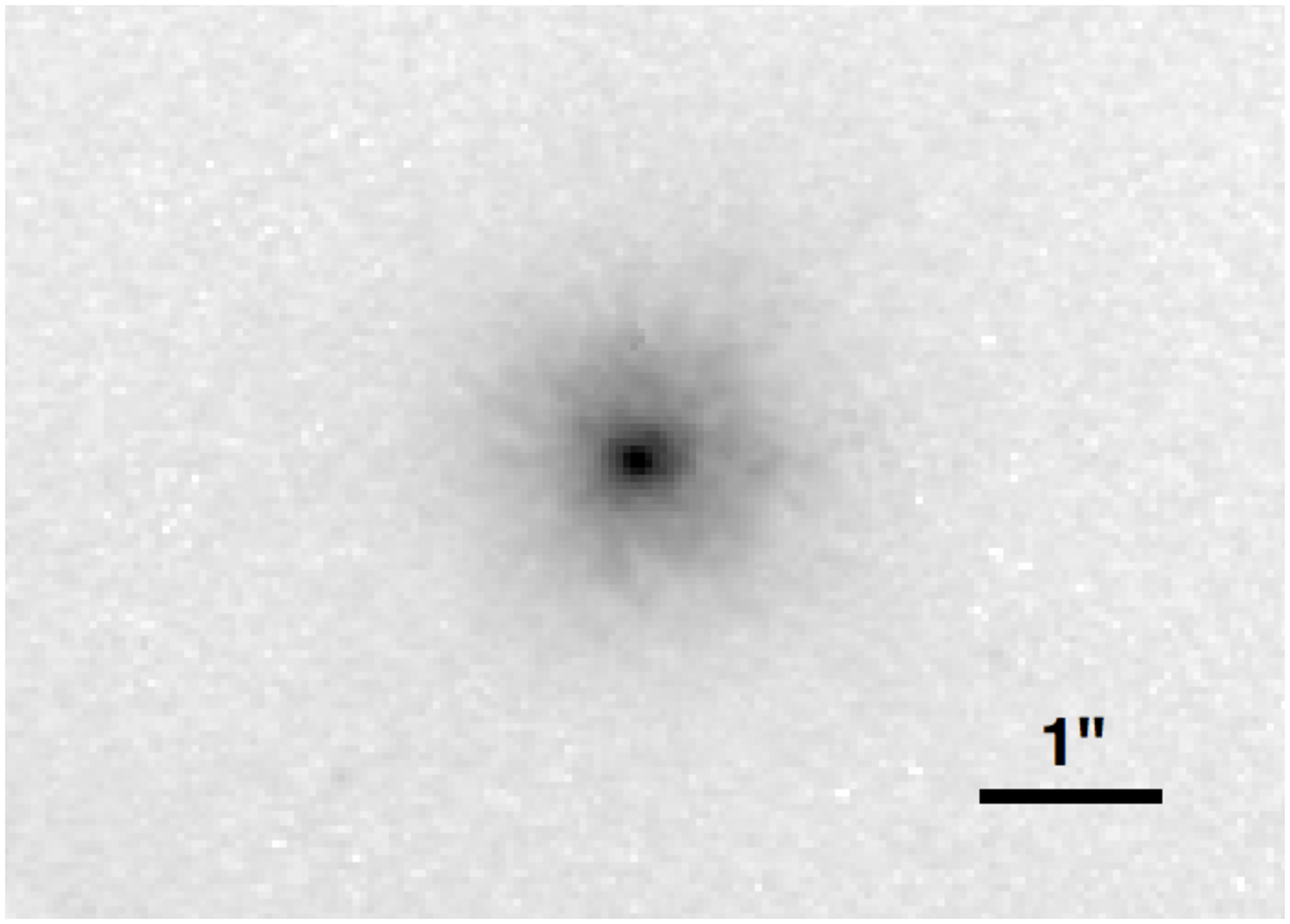} 
\caption{Adaptive optics image of HAT-P-14 in the $H$-band. The corrected PSF has a FWHM of 3 pixels (0.12\arcsec). No contaminating sources are detected to a limiting magnitude of H $\sim$ 17.1 mag at a distance of $>$ 0.48\arcsec\ from the centre of the corrected profile.  }
\label{AO}
\end{figure}

In order to search for additional close stars which may cause blends, we obtained high-resolution $H$- and $K$-band imaging with the near-infrared camera INGRID, fed by the adaptive optics system NAOMI on the 4.2-m William Herschel Telescope (WHT), see Figure \ref{AO}. Images were taken in two sky position angles (PA, 0\degr\ and 30\degr) in natural seeing of 0.5\arcsec, and the corrected PSFs have a FWHM of 0.12\arcsec and circular contours. No contaminating sources are detected to limiting magnitudes of $H$ $\sim$ 17.1 $\pm$ 0.25 and $K_{\rm s}$ $\sim$ 16.7 $\pm$ 0.25 at distances $>$ 4 $\times$ FWHM from the centre of the corrected profile. At 1.5 $\times$ FWHM the limiting magnitude is ~2 magnitudes brighter than this.

\section{Results} \label{Results}

\subsection{Stellar parameters}\label{stellar}
	
\begin{table}
\caption{Stellar parameters for HAT-P-14.}
\centering
\begin{tabular}{lc}
\hline
Parameter (Unit) & Value  \\
\hline
Photometric \& spatial properties: &\\
RA (J2000) & 17:20:27.88\\
DEC (J2000) & +38:14:31.7   \\
$V$(mag) & 9.98 $\pm$ 0.058 \\
$J$ (mag)&  9.094 $\pm$ 0.021  \\
$H$ (mag)&  8.927 $\pm$ 0.020  \\
$K_{s}$ (mag)& 8.851 $\pm$ 0.019\\
$\mu_{\rm RA}$ (mas year$^{-1}$) & 1.6 $\pm$ 0.7 \\
$\mu_{\rm DEC}$ (mas year$^{-1}$) & -4.7 $\pm$ 0.7 \\
$U$ (\kms) & 6.5 $\pm$ 1.0   \\
$V$ (\kms) & -10.6 $\pm$ 0.1  \\
$W$ (\kms) & -6.0 $\pm$ 0.8  \\
Galactic  longitude (deg) & 62.6 \\
Galactic latitude (deg) & 33.5 \\
\hline
Spectroscopic properties: & \\
\teff\ (K)    & 6583 $\pm$ 100 \\
\logg\ (cgs)     &  4.1 $\pm$ 0.12  \\
\mictrb\ (\kms)   & 0.85   \\
\mactrb\ (\kms) &  5.28  \\
\vsini\ (\kms)  & 8.4 $\pm$ 1.0 \\
{[M/H]}   & 0.08 $\pm$ 0.1 \\
\hline
Derived properties: & \\
\mstar\ (\msun) $^{a}$ & 1.30 $\pm$ 0.03  \\
\rstar\ (\rsun) $^b$ &  1.48 $\pm$ 0.05  \\
\rhostar\ (\rhosun) $^{c}$ &  0.398$^{+0.037}_{-0.034}$ \\
L$_{*}$ (L$_{\astrosun}$)$^{d}$ &  3.91$_{+0.54}^{-0.37}$ \\
Age (Gyr)$^{d}$ & 1.6$^{+0.4}_{-0.3}$ \\
Spectra type & F5V \\
\hline
\end{tabular}

\medskip
\flushleft
Note: The photometric and spatial properties are taken from or derived using the following sources: $V$ \citep[TASS,][]{TASS}, J, H, K$_{s}$ \citep[2MASS,][]{2MASS}, proper motions \citep[NOMAD,][]{NOMAD}, space velocities (calculated according to \citealt{Johnson87}, updated to J2000, for a right-hand coordinate system and corrected for solar motion using data from \citealt{Dehnen98}) and galactic co-ordinates (NED). \\
$^a$ derived from the empirical relationship of \citet{Torres10star}.\\
$^b$ determined using \mstar\ and \rhostar.\\
$^c$ found using the lightcurve geometry of WASP data.\\
$^d$ from stellar models
\label{sparams}
\end{table}

We derived the fundamental stellar parameters of HAT-P-14 using the following techniques, and the results are shown in Table \ref{sparams}. First, we performed a spectral analysis using SME \citep*[Spectroscopy Made Easy, see][]{VP96}, following the method of \citet{VF05} and using the Kurucz model atmospheres \citep{Kurucz84}. Individual NOT spectra were normalised and co-added to produce a single high signal-to-noise spectrum. The \halpha, Na {\sc i} D and Mg {\sc i} b lines were fitted simultaneously to determine the spectral parameters. This yielded the values \teff\ = 6583 $\pm$ 100 K and \logg$_{\rm spec}$ = 4.02 $\pm$ 0.1, which indicate that the star is of spectral type F5V \citep{Gray08}. Values for microturbulence (\mictrb\ = 0.85 \kms) and macroturbulence (\mactrb\ = 5.28 \kms) are taken from the prescription of \citet{VF05}. The projected stellar rotation velocity (\vsini) was determined to be \vsini\ = 8.4 $\pm$ 1.0 \kms\ and the metallically [M/H] = 0.08 $\pm$ 0.10. 

\begin{figure}	
\centering
\includegraphics[width=8cm, angle=-90]{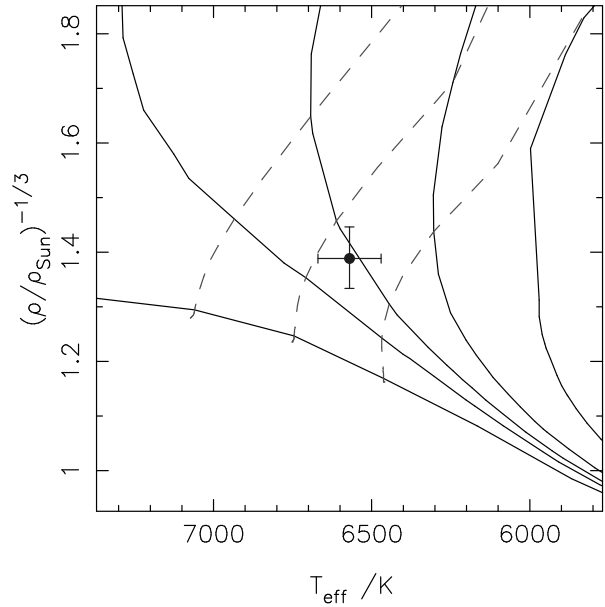} 
\caption{Theoretical stellar isochrones and mass tracks from \citet{Girardi00} plotted as the density proxy $R_{*}/M_{*}^{1/3}$ against effective temperature for [M/H] = 0.08. The solid lines indicate (from left to right) the ages 0.01, 1, 1.6, 2.5 and 5 Gyr and the dotted lines show masses of  1.5, 1.4 and 1.3 \msun. The position of HAT-P-14 is indicated within the error ranges. }
\label{iso}
\end{figure}

The mean stellar density ($\rhostar$), found from the light curve geometry, can be used as a luminosity indicator for stellar evolutionary models and often provides a stronger constraint than the value of \logg\ from spectral analysis \citep{Sozzetti07}. We used a Markov Chain Monte Carlo (MCMC) approach to globally model the photometric and radial velocity data (see Section \ref{Planet}) and obtained a mean stellar density of 0.398$^{+0.037}_{0.034}$ $\rho_{\astrosun}$. We then compared \teff, [M/H] and \rhostar\ with the theoretical stellar evolutionary models of \citet{Girardi00} to obtain the following stellar properties: \mstar\ = 1.36 $\pm$ 0.04 \msun, Age = 1.6 $^{+0.4}_{-0.3}$ Gyr,  L$_{*}$ = 3.91$_{-0.37}^{+0.54}$ L$_{\astrosun}$ and \logg$_{\rm iso}$ = 4.19 $\pm$ 0.03. The model isochrones are shown in Figure \ref{iso}.

The value of \logg\ obtained from the isochrone fit is somewhat larger than that found from spectral analysis, which was also noted by T10. We investigated the effect this higher \logg\ would have on the fit to the spectral line shapes by fixing \logg$_{\rm spec}$ to the isochrone value and reassessing the spectral analysis. We found that in order to retain a good fit to the Mg {\sc i} b lines, the magnesium abundance [Mg/H] must be reduced by 0.12 (in a similar fashion to WASP-1, see \citealt{Stempels07}) due to the inherent anti-correlation between [Mg/H] and \logg. The degeneracies between the spectroscopic parameters do not rule out the higher \logg\ value as long as [Mg/H] is allowed to be under-abundant, and we take the uncertainty in \logg\ to include both values. Altering \logg\ and [Mg/H] in this fashion had no significant effect on the derived effective temperature or metal abundance.  

\subsection{Planet parameters}\label{Planet}

To determine the properties of the planet HAT-P-14b, we simultaneously modelled the light curves and radial velocities with a global MCMC fit.  Details of this process are described in \citet{Cameron07} and \citet{Pollacco08}. The free parameters in the fit are: orbital period $P$; transit epoch $T_{0}$; transit duration $T_{\rm dur}$;  squared ratio of planet radius to star radius $R_{\rm P}^{2}$/R$_{*}^{2}$; impact parameter $b$; RV semi-amplitude $K$; Lagrangian elements $e\cos \omega$ and $e \sin \omega$, where $\omega$ is the longitude of periastron; and the systematic offset velocity $\gamma$. In this particular case, two systematic velocities were fit to allow for instrumental offsets between the SOPHIE and FIES datasets. These were found to be: $\gamma_{\rm SOPHIE}$ = -20.3834 $\pm$ 0.0027 \kms and $\gamma_{\rm FIES}$ = -20.4290 $\pm$ 0.0017 \kms. 

The chain length was 10,000 points and the resulting best-fit parameters and uncertainties are shown in Table \ref{pparams}. The photometric error bars were re-scaled to take into account any underestimation in the uncertainties so that $\chi^{2}_{red}$ = $\chi^{2}$/dof = 1 (dof = number of points - number of fitted parameters). We used the \citet{claret00, claret04} non-linear limb darkening coefficients for the appropriate stellar temperature and photometric passband appropriate for each light curve. 

In order to estimate the stellar mass and properties derived from it, we used the \citet{Torres10star} empirical calibrations from eclipsing binaries. This method, adapted by \citet{Enoch10}, computes the stellar mass as a function of \teff, $\log \rho_{*}$ and [M/H] and has the advantage of being independent of stellar models. We find \mstar = 1.30 $\pm$ 0.03 \msun, similar to the stellar mass derived from the isochrone analysis undertaken here (\mstar = 1.36 $\pm$ 0.04 \msun, Padova models) and in T10 (\mstar = 1.39 $\pm$ 0.05 \msun, YY models). 

We performed the MCMC fit and derived the following planetary parameters:  \mpl\ = 2.25 $\pm$ 0.06 \mj, \rpl\ =  1.19 $\pm$ 0.05 \rj\ and \rhopl\ = 1.33 $\pm$ 0.16 \rhoj. In addition, we analysed the data solely from T10 and found that the fitted parameters were completely consistent with their results. Finally, we analysed all the datasets simultaneously in order to optimise the best-fit solution using all the information available. The results are shown in Table \ref{pparams} and are in good agreement.

\subsubsection{Eccentricity}
In an initial fit to the SOPHIE and FIES RVs, the eccentricity was allowed to float, yielding $e$ = 0.054 $\pm$ 0.017. We analysed the datasets independently and found that a jitter term of 15 m s$^{-1}$ was required to be added in quadrature to the FIES RV uncertainties in order to obtain $\chi^{2}_{red}$ = 1. This may due to astrophysical noise, as also noted by T10, and/or instrumental systematic noise which has not been accounted for. The FIES RVs do not significantly detect an eccentric orbit, yielding e = 0.04 $\pm$ 0.03, but equally do not rule one out. 

In contrast, the SOPHIE RVs do not require any jitter term to reconcile the rms scatter with the measured uncertainties.  Fitting the SOPHIE RVs alone, a more significant detection of eccentricity was found, $e$ = 0.081 $\pm$ 0.022 with $\chi^{2}$ = 2.0 for 9 RV points and 4 fitted parameters ($K$, $e \cos \omega$, $e \sin \omega$, $\gamma_{\rm SOPHIE}$).  Fitting a purely circular orbit to the same data, we found $\chi^{2}_{\rm circ}$ = 22.6. Applying the Lucy-Sweeney test \citep*{Lucy71}, the eccentricity is detected at the 3$\sigma$ level.

A joint fit to both datasets reveals $e$ = 0.061$^{+0.022}_{-0.020}$, $\chi^{2}$ = 34.4 and $\chi^{2}_{\rm circ}$ = 49 for 33 RV points and 5 fitted parameters (including two systematic velocities, $\gamma_{\rm SOPHIE}$ and $\gamma_{\rm FIES}$), indicating the eccentricity is non-zero at the 2.7$\sigma$ level. We also apply the test to the T10 Keck RVs, including their 7.3 m s$^{-1}$ jitter term. We find the eccentricity is detected at the 4$\sigma$ level, $e$ = 0.104 $\pm$ 0.012 , with $\chi^{2}$ = 12.3, $\chi^{2}_{\rm circ}$ = 83.9, 14 RVs and 4 fitted parameters. By combining all three datasets, our analysis suggests that the orbit is not circular with a 6$\sigma$ confidence, $e$ = 0.095 $\pm$ 0.011 ($\chi^{2}$ = 55, $\chi^{2}_{\rm circ}$ = 133, 47 RVs and 6 fitted parameters (including three systematic velocities, $\gamma_{\rm SOPHIE}$, $\gamma_{\rm FIES}$ and $\gamma_{\rm Keck}$).

\begin{sidewaystable} 
\centering
\caption{System parameters for HAT-P-14} 
\begin{tabular}{lccc} 
\hline 
Parameter (Unit) & Value & \citet{Torres10planet} & Combined \\ 
\hline 
Photometric parameters: & & &\\
$P$ (d) &  4.627653 $\pm$ 0.000008 & 4.627669 $\pm$ 0.000005 &  4.627657 $\pm$ 0.000005\\
$T_{\rm 0}$ (HJD) & 2455301.03431 $\pm$ 0.00036 & 2455134.43854 $\pm$ 0.00026 & 2455134.43856 $\pm$ 0.00027\\
$T_{\rm dur}$ (d) & 0.0938 $\pm$ 0.0013 & 0.0912 $\pm$ 0.0017 &  0.0932 $\pm$ 0.0010\\
$\Delta F=R_{\rm P}^{2}$/R$_{*}^{2}$ & 0.00678 $\pm$ 0.00016 & 0.00648 $\pm$ 0.00024 & 0.00671 $\pm$ 0.00014 \\
$R_{\rm P}$/$R_{*}$ & 0.0824 $\pm$ 0.0010 & 0.0805 $\pm$ 0.0015 & 0.0819 $\pm$ 0.0009\\
$b$ & 0.915 $\pm$ 0.005 & 0.891$^{+0.007}_{-0.008}$ & 0.907 $\pm$ 0.004\\
$a/R_{*}$ & 8.60 $\pm$ 0.31 & 8.87 $\pm$ 0.29 & 8.49 $\pm$ 0.25  \\
$i$ (\degree) & 83.5 $\pm$ 0.4 & 83.5 $\pm$ 0.3 & 83.2 $\pm$ 0.2\\
\hline
Spectroscopic parameters: & & &\\
$K$ (\kms) & 0.2291 $\pm$ 0.0051 & 0.2190 $\pm$ 0.0033& 0.2224$\pm$ 0.00261\\
$e \cos \omega$ & -0.025 $\pm$ 0.01 & -0.009 $\pm$ 0.009& -0.014$\pm$ 0.006 \\
$e \sin \omega$ &  0.056  $\pm$ 0.025 & 0.106 $\pm$ 0.013 & 0.094 $\pm$ 0.012\\
$e$ & 0.061 $\pm$ 0.022 & 0.107 $\pm$ 0.013 & 0.095 $\pm$ 0.011\\
$\omega$ (\degree) & 114$^{+18}_{-11}$ & 94 $\pm$ 4 & 99 $\pm$ 4\\
\hline
Derived parameters: && &\\
$M_{\rm P}$ ($M_{\rm Jup}$) & 2.25 $\pm$ 0.06 & 2.23 $\pm$ 0.06& 2.20 $\pm$ 0.04\\
$R_{\rm P}$ ($R_{\rm Jup}$) & 1.19 $\pm$ 0.05 & 1.15 $\pm$ 0.05 & 1.20 $\pm$ 0.04\\
$\rho_{\rm P}$ ($\rho_{\rm J}$) & 1.33 $\pm$ 0.16 & 1.47 $\pm$ 0.20 & 1.28 $\pm$ 0.12 \\
$\log g_{\rm P}$ (cgs) & 3.56 $\pm$ 0.03 & 3.62 $\pm$ 0.04 & 3.54 $\pm$ 0.03  \\
$a$ (AU)  & 0.0591 $\pm$ 0.0004 &0.0606 $\pm$ 0.0007 & 0.0594 $\pm$ 0.0004 \\
$T_{\rm eq, A=0}$ (K) & 1585 $\pm$ 35 & 1570 $\pm$ 34& 1597 $\pm$ 29 \\
\hline 
\label{pparams} 
\end{tabular} 
\end{sidewaystable} 

\section{Conclusions} \label{Conc}
We report the independent discovery and follow-up observations of the transiting planet HAT-P-14b found by T10. We have analysed the follow-up data used to secure the detection, and derived an independent set of planetary parameters which are consistent with those found by T10. We provide additional evidence, through AO observations and additional bisector span measurements, that the planetary signal is not caused by astronomical false positives.

We find the properties of the parent star to be in good agreement with those found by T10, which is reassuring given the different techniques used in their derivation. The star is a hot F5 dwarf with \teff\ = 6583 $\pm$ 100 K and appears to be relatively young at 1.6$^{+0.4}_{-0.3}$ Gyr, with a slightly super-solar metallicity [M/H] = 0.08. Using the empirical calibrations of \citet{Torres10star} we determine the stellar mass to be 1.30 $\pm$ 0.03 \msun, which agrees well with the values we derive from theoretical stellar models.

By combining data from both groups, we are able to refine the orbital ephemeris $P$ = 4.627652 $\pm$ 0.000004 d,  $T_{\rm 0}$ = 2455134.43823 $\pm$ 0.00022 d, and planetary parameters. HAT-P-14b has \rpl\ = 1.20 $\pm$ 0.05 \rj, \mpl\ = 2.20 $\pm$ 0.04 \mj\ and \rhopl\ = 1.28 $\pm$ 0.12 \rhoj. It appears to blur the bi-modal distribution noted by \citet{Bakos10b} in which massive (\mpl $\sim$ 2 \mj) planets appear either inflated (\rpl $\sim$ 1.3 \rj) or compact (\rpl $\sim$ 1 \rj). We find the orbit to be significantly eccentric, $e$ =  0.095 $\pm$ 0.011, which hints at the possibility of another body in the system. 

As noted by T10, the system promises excellent opportunities for further follow-up and characterisation, in particular due to the brightness of the host star, $V_{\rm mag}$ = 9.98. The star is relatively fast rotating for a planet host, \vsini\ = 8.4 $\pm$ 1.0 \kms, making it a good candidate for the detection of the Rossiter-McLaughlin effect \citep[RM,][]{Rossiter24,McLaughlin24}. However, due to the very high impact parameter, b = 0.905 $\pm$ 0.005, the RM amplitude is only expected to be $\sim$ 20 m/s. Despite this, it is still an attractive target as \citet{Winn10} and \citet{Schlaufman10} have shown that misaligned planets appear to preferentially orbit stars with high \teff\ and \mstar\ such as HAT-P-14.  

The high temperature of the host star places HAT-P-14b in the highly-irradiated `pM' category of \citet{Fortney08} and is a good candidate for investigating temperature inversions and day/night contrasts. An observation of the secondary transit would be particularly interesting to further constrain the low eccentricity. However, the orbital geometry of the system means that HAT-P-14b is on the borderline of having no secondary eclipse. The results from the joint analysis predict that it will not occur with more than 3$\sigma$ confidence. However, the fit to the data presented in this paper alone suggests that the secondary eclipse could be grazing. Since these opposing secondary eclipse predictions arise from extremely small differences in orbital eccentricity, whether or not we observe such an transit will provide further constraints on the system geometry. HAT-P-14 thus provides excellent opportunities for future investigations and informative characterisation.   

\section*{Acknowledgments}

The SuperWASP Consortium consists of astronomers primarily from QueenÕs University Belfast, St Andrews, Keele, Leicester, The Open University, Isaac Newton Group La Palma and Instituto de Astrofõsica de Canarias.  The SuperWASP-N camera is hosted by the Issac Newton Group on La Palma and we are grateful for their continuing support and assistance. Funding for WASP comes from consortium universities and from the UK's Science and Technology Facilities Council. Based on observations made at Observatoire de Haute Provence (CNRS), France and at the Nordic Optical Telescope, operated on the island of La Palma jointly by Denmark, Finland, Iceland, Norway, and Sweden, in the Spanish Observatorio del Roque de los Muchachos of the Instituto de Astrofisica de Canarias. The Liverpool Telescope is operated on the island of La Palma by Liverpool John Moores University in the Spanish Observatorio del Roque de los Muchachos of the Instituto de Astrofisica de Canarias with financial support from the UK Science and Technology Facilities Council. We thank Tom Marsh for the use of the ULTRACAM pipeline. FPK is grateful to AWE Aldermaston for the award of a William Penny Fellowship. EKS would like to thank R. M. Crockett for his thorough proof-reading and excellent image-making.  

\bibliographystyle{aa}
\bibliography{bib.bib}

\end{document}